\documentclass[prd,preprint,superscriptaddress,showpacs,byrevtex]{revtex4}
\usepackage{epsfig}

\newcommand{\be}{\begin{equation}}
\newcommand{\ee}{\end{equation}}
\newcommand{\bea}{\begin{eqnarray}}
\newcommand{\eea}{\end{eqnarray}}

\begin{document}

\begin{flushright}
YITP-SB-07-17
\end{flushright}

\title{ Path Integration in QCD with Arbitrary Space-Dependent
Static Color Potential }

\author{Gouranga C. Nayak} \email{nayak@max2.physics.sunysb.edu}

\affiliation{ C. N. Yang Institute for Theoretical Physics, Stony
Brook University, SUNY, Stony Brook, NY 11794-3840, USA }

\date{\today}

\begin{abstract}

We perform the path integral
for a quark (antiquark) in the presence of an external background chromo-electric
SU(3) gauge field $E^a(x^1=x)$ with arbitrary color index $a$=1,2,...8 and
obtain an exact non-perturbative expression for the generating functional.
The only nonzero field strength component considered is $E^a=F^a_{01}$
which is allowed to depend on a single spatial coordinate $x^1=x$. We show that
such a path integration is possible even if one can not solve the Dirac equation
in the presence of arbitrary space-dependent potential. This result crucially
depends on the validity of the shift conjecture which has not yet been
established. It may be possible to further explore this path integral technique
to study non-perturbative bound state formation.

\end{abstract}
\maketitle

\newpage

\section{Introduction}

In QED and in Newtonian gravity the potential
(due to charge and mass respectively) is inversely proportional to distance.
However, the exact form of the potential due to color charge is not
known in QCD. This is a fundamental problem in physics. Quarks
and gluons which carry color charges are confined inside hadrons.

Lattice QCD simulation attempts to compute the
form of the static potential between quark and antiquark inside quarkonium
\cite{wilson,latt}. The static potential inside heavy quarkonia can be approximated
to be a Coulomb plus linear form \cite{latt1}. However, the exact form of the
potential inside light $q\bar q$ mesons and inside other hadrons
is expected to be quite complicated. The long distance confinement physics
is non-perturbative, and one can not perform
path integrations analytically.
On the other hand the lattice QCD method implements
extensive numerical simulations and hence the physical picture is not
as clear as it could be if one had explicit analytic results.
From this point of view path integration in the presence of arbitrary
space-dependent static color potential would be desirable.

In this paper we present an analytical method to perform
the path integral for a quark (antiquark)
in the presence of arbitrary space-dependent
static color potential $A^a_0(x) (=-\int dx E^a(x)$)
with arbitrary color index $a$=1,2,..8 in SU(3).
We show that such a path integration is
possible even if one can not solve the Dirac equation in the presence
of arbitrary space-dependent potential. This result crucially
depends on the validity of the shift conjecture which has not yet been
established. Unlike QED, the above potential
in QCD is motivated because the color quark-antiquark potential
may be one dimensional.
In particular we obtain the following exact non-perturbative expression
for the generating functional for a quark (antiquark) in the presence of
arbitrary space-dependent static color potential $A^a_0(x)=-\int dx E^a(x)$:
\bea
&& \frac{Z[A]}{Z[0]}=\frac{\int [d\bar{\psi }][d\psi ] e^{i\int d^4x \bar{\psi}^j(x)(\delta_{jk}{{\hat p}\!\!\!\slash} ~-gT^a_{jk}{A\!\!\!\slash}^a -\delta_{jk}m) \psi^k(x)}}{\int [d\bar{\psi }][d\psi ] e^{i\int d^4x \bar{\psi}^j(x)({{\hat p}\!\!\!\slash } ~-m) \psi^k(x)}} =  \nonumber \\
&&\exp \big \{\frac{i }{8\pi^2}~\sum_{j=1}^3 \int dt \int d^3x \int_0^\infty \frac{ds}{s^2}e^{-is(m^2-i\epsilon)}
~[
g\Lambda_j(x) ~{\rm coth} (gs \Lambda_j(x))
-\frac{1}{3}sg^2\Lambda^2_j(x)
-\frac{1}{s}
]\big \}. \nonumber \\
\label{n}
\eea
The gauge invariant $\Lambda_j(x)$'s are given by
\bea
&& \Lambda_1(x)=\sqrt{\frac{C_1(x)}{3}} {\rm cos} \theta(x),~~~
\Lambda_{2,3}(x)=\sqrt{\frac{C_1(x)}{3}} {\rm cos}(\frac{2\pi}{3}\pm \theta(x)),~~~
{\rm cos}^23 \theta(x)=\frac{3C_2(x)}{C_1^3(x)} \nonumber \\
\label{lambc}
\eea
where $C_1(x)=[E^a(x)E^a(x)]$ and
$C_2(x)=[d_{abc}E^a(x)E^b(x)E^c(x)]^2$
are two independent space-dependent
casimir/gauge invariants in SU(3). The color
indices $a,b,c$=1,2,...8.

We will present a derivation of eq. (\ref{n}) in this paper.
It may be possible to further explore this path integral technique
to study non-perturbative bound state formation, for example, by studying
various non-perturbative correlation functions.

\section{Path Integration With Arbitrary space-Dependent
Static Color Potential }

The Lagrangian density for quark in a classical chromofield $A_{\mu}^a$
is given by
\be
{\cal{L}}~
=~\bar{\psi}^j~ [(\delta_{jk}~\hat{{p}\!\!\!\slash} ~-~gT^a_{jk}{{A}\!\!\!\slash}^a) -m\delta_{jk}]
~\psi^k ~=~\bar{\psi}^j~ M_{jk}[A] ~\psi^k
\label{laq}
\ee
where $\hat{p}_\mu =\frac{1}{i} \frac{\partial}{\partial x^\mu}$
is the momentum operator and
$T^a_{jk}$ is the generator in the fundamental representation of gauge group SU(3) with
$a$=1,2...8 and $j,k$~=~1,2,3. The generating functional is given by
\bea
\frac{Z[A]}{Z[0]}=\frac{\int [d\bar{\psi}][d\psi] e^{i\int d^4x~\bar{\psi}^j M_{jk}[A] \psi^k}}{
\int [d\bar{\psi}][d\psi] e^{i\int d^4x~\bar{\psi}^j M_{jk}[0]~\psi^k}} =
{\rm \frac{Det[M[A]]}{Det[M[0]]} }=e^{iW}.
\label{gen}
\eea
This gives
\bea
W= -\frac{i}{2}{\rm Tr~ln}
[(\delta_{jk} \hat{{p}\!\!\!\slash} -gT^a_{jk}{{A}\!\!\!\slash}^a)^2 -m^2\delta_{jk}]
+\frac{i}{2}{\rm Tr~ln} [\delta_{jk} (\hat{{p}}^2  -m^2)]
\label{efg3}
\eea
where
\bea
{\rm Tr {\cal O}}={\rm tr}_{\rm Dirac} {\rm tr}_{\rm color} \int dt
\int dx \int dy \int dz <t|<x|<y|<z| {\cal O} |z> |y> |x> |t>.
\eea

Eq. (\ref{efg3}) can be written as
\bea
W=\frac{i}{2} \int_0^\infty \frac{ds}{s} {\rm Tr}
[e^{is[(\delta_{jk} \hat{{p}} -gT^a_{jk}{{A}}^a)^2+ \frac{g}{2}\sigma^{\mu \nu}T^a_{jk}F^a_{ \mu \nu} -\delta_{jk}m^2+i\epsilon]} -e^{is[\delta_{jk}(\hat{p}^2  -m^2)+i\epsilon]}].
\label{3c}
\eea

\subsection{ Suitable Gauge Choice for Evaluation of this Path Integral }

A physical quantity constructed from the
arbitrary space-dependent static color potential
\bea
A_\mu^a(x) = -\delta_{\mu 0} \int dx E^a(x)
\label{pot}
\eea
may be expressed in terms of two space-dependent casimir/gauge invariants in SU(3)
\bea
C_1(x)= [E^a(x)E^a(x)],~~~{\rm and},~~~C_2(x)=[d_{abc}E^a(x)E^b(x)E^c(x)]^2.
\label{cas}
\eea
However, it is not possible to perform the above path integration
by directly using eq. (\ref{pot}) in eq. (\ref{3c}).
For this reason we use the gauge invariance argument.
Since $A_\mu^a(x)$ is not gauge invariant we can work in a
different gauge to obtain the same $E^a(x)$. For example, if we choose
\bea
A^a_\mu(t,x) =-\delta_{\mu 1} E^a(x)t
\label{7c}
\eea
then we reproduce the same chromo-electric field $E^a(x)$ as
in eq. (\ref{pot}). Note that both the equations (\ref{pot}) and
(\ref{7c}) are related by a gauge transformation
and reproduce the same $E^a(x)$ but eq. (\ref{7c})
is preferred to do path integration because the 't' variable
will allow us to cast this as a harmonic oscillator problem.
This is not possible by eq. (\ref{pot}). Hence we can use eq.
(\ref{7c}) instead of eq. (\ref{pot}) to perform path integration
in the presence of arbitrary space-dependent static color
potential.

Using eq. (\ref{7c}) in (\ref{3c}) we find
\bea
&& W
=\frac{i}{2} ~{\rm tr}_{\rm Dirac} {\rm tr}_{\rm color} \int_0^\infty \frac{ds}{s} \int dt \int dx \int dy \int dz <t| <x| <y| <z| \nonumber \\
&&
[e^{is [-(\delta_{jk}\hat{p}_1+gT^a_{jk}E(x) t)^2+\delta_{jk}(\hat{p}_0^2- \hat{p}_y^2-\hat{p}_z^2)+ ig\gamma^0 \gamma^1 T^a_{jk} E^a(x)-\delta_{jk}m^2+i\epsilon]} -e^{is(\delta_{jk}(\hat{p}^2-m^2)+i\epsilon)}] \nonumber \\
&& |z>|y>|x>|t>.
\label{10cfb}
\eea
Inserting complete set of $|p>$ states ($\int dp~|p><p|=1$) as
appropriate we obtain
(we use the normalization $<q|p>=\frac{1}{\sqrt{2\pi}} e^{iqp}$)
\bea
&& W= \frac{i}{2(2\pi)^2} {\rm Tr_{Dirac}} {\rm Tr_{color}} [
\int_0^\infty \frac{ds}{s} \int_{-\infty}^{+\infty} dy \int_{-\infty}^{+\infty} dz \int dp_y \int dp_z e^{-is(p_y^2+p_z^2+m^2-i\epsilon)} \int dx \int dt \nonumber \\
&&~[
<x| <t| e^{is[ -(\frac{\delta_{jk}}{i}\frac{d}{dx}+gT^a_{jk}E^a(x)t)^2+\delta_{jk} \hat{p}_0^2+ig\gamma^0 \gamma^1 T^a_{jk}E^a(x)]} |t>|x>-
\frac{\delta_{jk}}{s} ]].
\label{5cs}
\eea
To perform the Dirac trace we use the eigenvalues of the Dirac matrix
\bea
(\gamma^0 \gamma^1)_{\rm eigenvalues} =(\lambda_1, \lambda_2, \lambda_3, \lambda_4) =(1,1,-1,-1)
\label{eigend}
\eea
and find
\bea
&& W= \frac{i}{2(2\pi)^2} \sum_{l=1}^4
\int_0^\infty \frac{ds}{s} \int_{-\infty}^{+\infty} dy \int_{-\infty}^{+\infty} dz \int dp_y \int dp_z e^{-is(p_y^2+p_z^2+m^2-i\epsilon)} \int dx \int dt \nonumber \\
&&~[ F_l-\frac{3}{s}],
\label{15cs}
\eea
where
\bea
F_l={\rm tr}_{\rm color} \int_{-\infty}^{+\infty} dx <x| \int_{-\infty}^{+\infty} dt <t| e^{is[ -(\frac{\delta_{jk}}{i}\frac{d}{dx}+gT^a_{jk}E^a(x)t)^2+ \delta_{jk}\hat{p}_0^2+ig\lambda_l T^a_{jk}E^a(x)]} |t>|x>.
\label{j1}
\eea
In the matrix notation we write the above equation as
\bea
F_l={\rm tr}_{\rm color} [\int_{-\infty}^{+\infty} dx <x| \int_{-\infty}^{+\infty} dt <t| e^{is[ -(\frac{1}{i}\frac{d}{dx}+gM(x)t)^2+ \hat{p}_0^2+ig\lambda_l M(x)]} |t>|x>]_{jk}
\label{i1}
\eea
where
\bea
M_{jk}(x)=T^a_{jk}E^a(x).
\label{mt}
\eea

\subsection{ Trouble with the Color Trace Evaluation in this Path Integral }

Unlike the constant chromo-electric field case \cite{nayak1}
it is not straight forward to take the color trace by diagonalizing
the color matrix $T^a_{jk}E^a(x)$ by an orthogonal matrix
$U_{jk}(x)$. This is because the
orthogonal matrix $U_{jk}(x)$ is $x$ dependent and does not
commute with $\frac{d}{dx}$. This was not a problem for
constant chromo-electric field $E^a$ case \cite{nayak1} because
the orthogonal matrix $U_{jk}$ was space-time independent.
This trouble was also not there in Schwinger mechanism study
in QED \cite{nayak2} because there are no colors in QED.
This implies that as long as the derivative operator $\frac{d}{dx}$
is present in the expression for $F_l$ we can not take the color trace
by diagonalizing the space-dependent color matrix
$M_{jk}(x)=T^a_{jk}E^a(x)$ by an orthogonal matrix $U_{jk}(x)$. We
will come back to this color trace issue later.

We note that since $M(x)$ and $t$ in eq. (\ref{i1}) commutes with each other
one can not change the integration
variable from $t$ to $t'$ via: $t=t'-\frac{1}{igM(x)}\frac{d}{dx}$
in color space.
This is because $M(x)$ and $\frac{d}{dx}$ do not commute with each other.
To deal with the
$x$-dependent color matrix $M_{jk}(x)$ and $\delta_{jk} \frac{d}{dx}$
we use the similarity transformation of $t$ in the color space
\bea
[t \pm \frac{1}{i gM(x)} \frac{d}{dx}]_{jk}=
[e^{\pm \frac{1}{igM(x)}\frac{d}{dx}\frac{d}{dt}}
t e^{\mp \frac{1}{igM(x)}\frac{d}{dx}\frac{d}{dt}}]_{jk}
\label{sim1}
\eea
which gives
\bea
[gM(x)t + \frac{1}{i} \frac{d}{dx}]_{jk}=
[ e^{ \frac{1}{i gM(x)}\frac{d}{dx}\frac{d}{dt}}
[e^{-\frac{1}{i gM(x)}\frac{d}{dx}\frac{d}{dt}} gM(x)
e^{\frac{1}{i gM(x)}\frac{d}{dx}\frac{d}{dt}}
t]
e^{- \frac{1}{i gM(x)}\frac{d}{dx}\frac{d}{dt}}]_{jk}.
\label{sim2}
\eea
Using this in eq. (\ref{i1}) we find
\bea
&& F_l={\rm tr}_{\rm color} [\int_{-\infty}^{+\infty} dx <x| \int_{-\infty}^{+\infty} dt <t|
e^{ \frac{1}{i gM(x)}\frac{d}{dx}\frac{d}{dt}} \nonumber \\
 && e^{is[ -
[e^{- \frac{1}{i gM(x)}\frac{d}{dx}\frac{d}{dt}} gM(x)
e^{ \frac{1}{i gM(x)}\frac{d}{dx}\frac{d}{dt}}
t]^2
+ \hat{p}_0^2+ig\lambda_l
e^{- \frac{1}{i gM(x)}\frac{d}{dx}\frac{d}{dt}}
M(x)
e^{ \frac{1}{i gM(x)}\frac{d}{dx}\frac{d}{dt}}
]} \nonumber \\
&& e^{- \frac{1}{i gM(x)}\frac{d}{dx}\frac{d}{dt}}|t>|x>]_{jk}.
\label{ij1}
\eea

It can be seen that eq. (\ref{i1}) contains the product $t M(x)$ which 
satisfies 
\bea
t ~M(x) = M(x)~ t
\label{tm}
\eea
because $M(x)$ is independent of $t$. However, when we change the variable
\bea
t=t'-\frac{1}{igM(x)}\frac{d}{dx}
\label{tt1}
\eea
we find
\bea
t ~M(x) \neq M(x)~t.
\label{tnm}
\eea
Hence one can not change the variable $t$ by eq. (\ref{tt1}) in (\ref{i1}).
However, in eq. (\ref{ij1}) there is no product $t~M(x)$ present. 
Hence we can change the integration variable
$t$ to $t'$ via:
$t=t'-\frac{1}{igM(x)}\frac{d}{dx}$ in the color space in eq. (\ref{ij1}).
Since the $t$ integration limit is from $-\infty$ to $+\infty$, the $t'$
integration limit also remains from $-\infty$ to $+\infty$. Hence we
find from eq. (\ref{ij1})
\bea
&& F_l={\rm tr}_{\rm color} [\int_{-\infty}^{+\infty} dx <x| \int_{-\infty}^{+\infty} dt <t-\frac{1}{igM(x)}\frac{d}{dx}|
e^{ \frac{1}{i gM(x)}\frac{d}{dx}\frac{d}{dt}} \nonumber \\
 && e^{is[ -
[e^{- \frac{1}{i gM(x)}\frac{d}{dx}\frac{d}{dt}} gM(x)
e^{ \frac{1}{i gM(x)}\frac{d}{dx}\frac{d}{dt}}
(t-\frac{1}{igM(x)}\frac{d}{dx})]^2
+ \hat{p}_0^2+ig\lambda_l
e^{- \frac{1}{i gM(x)}\frac{d}{dx}\frac{d}{dt}}
M(x)
e^{ \frac{1}{i gM(x)}\frac{d}{dx}\frac{d}{dt}}
]} \nonumber \\
&& e^{- \frac{1}{i gM(x)}\frac{d}{dx}\frac{d}{dt}}
|t-\frac{1}{igM(x)}\frac{d}{dx}>|x>]_{jk}.
\label{ij2}
\eea
Using the similarity transformation of
$[ t-\frac{1}{igM(x)}\frac{d}{dx}]_{jk}$ from eq. (\ref{sim1})
we find from the above equation
\bea
&&F_l={\rm tr}_{\rm color} [\int_{-\infty}^{+\infty} dx <x| \int_{-\infty}^{+\infty} dt <t-\frac{1}{igM(x)}\frac{d}{dx}|
e^{ \frac{1}{i gM(x)}\frac{d}{dx}\frac{d}{dt}} \nonumber \\
 && e^{is[ -
[e^{- \frac{1}{i gM(x)}\frac{d}{dx}\frac{d}{dt}} gM(x)
e^{ \frac{1}{i gM(x)}\frac{d}{dx}\frac{d}{dt}}
e^{- \frac{1}{i gM(x)}\frac{d}{dx}\frac{d}{dt}}
t e^{ \frac{1}{i gM(x)}\frac{d}{dx}\frac{d}{dt}}]^2
+ \hat{p}_0^2+ig\lambda_l
e^{- \frac{1}{i gM(x)}\frac{d}{dx}\frac{d}{dt}}
M(x)
e^{ \frac{1}{i gM(x)}\frac{d}{dx}\frac{d}{dt}}
]} \nonumber \\
&& e^{- \frac{1}{i gM(x)}\frac{d}{dx}\frac{d}{dt}}
|t-\frac{1}{igM(x)}\frac{d}{dx}>|x>]_{jk}
\eea
which gives
\bea
&&F_l={\rm tr}_{\rm color} [\int_{-\infty}^{+\infty} dx <x| \int_{-\infty}^{+\infty} dt <t-\frac{1}{igM(x)}\frac{d}{dx}|
e^{is[ -g^2M^2(x) t^2+\hat{p}_0^2+ig\lambda_l M(x)]}
|t-\frac{1}{igM(x)}\frac{d}{dx}> \nonumber \\
&& |x>]_{jk}.
\label{i2sa}
\eea

It has to be remembered that eq. (\ref{i2sa}) is not valid if
the $t$ integration limit was finite \cite{shift}.
Hence we must perform the $t$ integration from
$-\infty$ to $+\infty$ in $F_l$ which we will do later in the derivation.

Inserting complete set of $|p_0>$ states we find
\bea
&&F_l ={\rm tr}_{\rm color} [\int_{-\infty}^{+\infty}
dx <x| \int_{-\infty}^{+\infty} dt \int dp'_0 \int dp''_0
<t-\frac{1}{igM(x)}\frac{d}{dx}|p'_0> \nonumber \\
&& <p'_0| e^{is[ -g^2M^2(x) t^2+\hat{p}_0^2+ig\lambda_l M(x)]} |p''_0>
<p''_0|t-\frac{1}{igM(x)}\frac{d}{dx}> |x>]_{jk} \nonumber \\
&& =\frac{1}{2\pi} {\rm tr}_{\rm color} [\int_{-\infty}^{+\infty}
dx <x| \int_{-\infty}^{+\infty} dt \int dp'_0 \int dp''_0
e^{itp'_0}e^{-\frac{p'_0}{gM(x)}\frac{d}{dx}} <p'_0|
e^{is[ -g^2M^2(x) t^2+\hat{p}_0^2+ig\lambda_l M(x)]} |p''_0> \nonumber \\
&& e^{-itp''_0}e^{\frac{p''_0}{gM(x)}\frac{d}{dx}} |x>]_{jk}.
\label{i2sb}
\eea
The above equation contains $\frac{d}{dx}$ and hence
we can not perform the color trace by diagonalizing the
color matrix $M_{jk}(x)$ by an orthogonal matrix $U_{jk}(x)$.

Inserting complete sets of states as appropriate we find
\bea
&& F_l=
\frac{1}{2\pi} {\rm tr}_{\rm color} [\int_{-\infty}^{+\infty} dx \int dx'
\int dx'' \int dp_x \int dp'_x \int dp''_x \int dp'''_x
<x|p_x> \int_{-\infty}^{+\infty} dt \int dt' \int dt'' \nonumber \\
&& \int dp'_0 \int dp''_0 e^{itp'_0}<p_x| e^{-\frac{1}{gM(x)}\frac{d}{dx}p'_0}|p'_x> <p'_0|t'> <p'_x|x'><x'|<t'|e^{is[ -g^2M^2(x)t^2+\hat{p}_0^2+ig\lambda_l M(x)]} \nonumber \\
&& |t''> |x''><x''|p''_x><t''|p''_0><p''_x|e^{\frac{1}{gM(x)}\frac{d}{dx}p''_0}|p'''_x>e^{-itp''_0}<p'''_x|x>]_{jk}.
\eea
Since the matrix $[<p''_x|e^{\frac{1}{gM(x)}\frac{d}{dx}p''_0}|p'''_x>]_{jk}$ is inside the trace and is independent of
$\frac{d}{dx}$ (it depends on c-numbers $p''_x$ and $p'''_x$), we can move it
to the left. We find
\bea
&&F_l=\frac{1}{(2\pi)^4} {\rm tr}_{\rm color} [\int_{-\infty}^{+\infty} dx
\int dx' \int dp_x \int dp'_x \int dp''_x \int dp'''_x
e^{ixp_x} \int_{-\infty}^{+\infty} dt \int dt' \int dt'' \nonumber \\
&& \int dp'_0 \int dp''_0 e^{itp'_0} <p''_x|e^{\frac{1}{gM(x)}\frac{d}{dx}p''_0}|p'''_x><p_x|e^{-\frac{1}{gM(x)}\frac{d}{dx}p'_0}|p'_x> e^{-ip'_0t'} e^{-ix'p'_x} \nonumber \\
&& <t'|e^{is[ -g^2M^2(x')t^2+\hat{p}_0^2+ig\lambda_l M(x')]}|t''>
e^{ix'p''_x}e^{ip''_0t''}e^{-itp''_0}e^{-ixp'''_x}]_{jk}.
\label{i2ss2}
\eea
The matrix
$[<p''_x|~e^{\frac{1}{gM(x)}\frac{d}{dx}p''_0}~|p'''_x>]_{jk}$ is also
independent of $x$. This can be shown as follows
\bea
&& <p''_x|f(x) \frac{d}{dx}|p'''_x>=\int dx' \int dx'' \int dp''''_x
<p''_x|x'> <x'|f(x)|x''> <x''|p''''_x> <p''''_x|\frac{d}{dx} |p'''_x> \nonumber \\
&& = \int dx' e^{-ix'(p''_x-p'''_x)} f(x')i p'''_x
\label{indt}
\eea
which is independent of $x$ and $\frac{d}{dx}$. Hence all the expressions in eq. (\ref{i2ss2})
are independent of $x$ except $e^{ix(p_x-p'''_x)}$. This implies that we can easily perform
the $x$ integration in eq. (\ref{i2ss2})
(by using $\int_{-\infty}^{+\infty} dx ~e^{ix(p_x-p'''_x)} = 2\pi \delta_(p_x-p'''_x))$ to obtain
\bea
&& F_l=\frac{1}{(2\pi)^3} {\rm tr}_{\rm color} [\int dx' \int dp_x \int dp'_x \int dp''_x
\int_{-\infty}^{+\infty} dt \int dt' \int dt'' \int dp'_0 \int dp''_0 \nonumber \\
&& e^{itp'_0} <p''_x| e^{\frac{1}{gM(x)}\frac{d}{dx}p''_0}|p_x><p_x|e^{-\frac{1}{gM(x)}\frac{d}{dx}p'_0}|p'_x> e^{-ip'_0t'} \nonumber \\
&& e^{-ix'p'_x}<t'|e^{is[ -g^2M^2(x')t^2+\hat{p}_0^2+ig\lambda_l M(x')]}|t''>e^{ix'p''_x}e^{ip''_0t''}e^{-itp''_0}]_{jk}.
\label{pq1}
\eea
As advocated earlier in eq. (\ref{i2sa}) we must integrate
over $t$ from $-\infty$ to $+\infty$ in $F_l$. Since
$<t'|e^{is[ -g^2M^2(x')t^2+\hat{p}_0^2+ig\lambda_l M(x')]}|t''>$ is independent of
$t$ variable (it depends on $t'$ and $t''$ variables) all the expressions in
eq. (\ref{pq1}) are independent of $t$ except $e^{it(p'_0-p''_0)}$. Hence we can easily
integrate over $t$ (by using $\int_{-\infty}^{+\infty} dt ~e^{it(p'_0-p''_0)}=
2\pi \delta(p'_0-p''_0))$ to obtain
\bea
&& F_l=\frac{1}{(2\pi)^2} {\rm tr}_{\rm color} [\int dx' \int dp_x \int dp'_x \int dp''_x
\int dt' \int dt'' \int dp'_0 \nonumber \\
&& <p''_x| e^{\frac{1}{gM(x)}\frac{d}{dx}p'_0}|p_x><p_x|e^{-\frac{1}{gM(x)}\frac{d}{dx}p'_0}|p'_x> e^{-ip'_0t'} \nonumber \\
&& e^{-ix'p'_x}<t'|e^{is[ -g^2M^2(x')t^2+\hat{p}_0^2+ig\lambda_l M(x')]}|t''>e^{ix'p''_x}e^{ip'_0t''}]_{jk}.
\label{pq}
\eea
Since $\int dp_x |p_x><p_x|=1$ we find from the above equation
\bea
&& F_l
=\frac{1}{(2\pi)^2} {\rm tr}_{\rm color} [\int dx' \int dp'_x
 \int dt' \int dt'' \int dp'_0 e^{-it' p'_0} \nonumber \\
&& <t' |e^{is[ -g^2M^2(x')t^2+\hat{p}_0^2+ig\lambda_l M(x')]}|t''> e^{it'' p'_0}]_{jk}.
\label{i2ss4}
\eea

\subsection{ Evaluation of the Color Trace }

As the above equation does not contain any derivative $\frac{d}{dx}$ operator
we can perform the color trace by diagonalizing the space-dependent
color matrix $M_{jk}(x)$ by a space-dependent orthogonal matrix $U_{jk}(x)$.
In the fundamental representation the matrix $M_{jk}(x)$ has three eigenvalues:
\bea
[M_{jk}(x)]_{\rm eigenvalues}=[T^a_{jk}E^a(x)]_{\rm eigenvalues} =(\Lambda_1(x), ~\Lambda_2(x), ~\Lambda_3(x)).
\label{eigenc}
\eea
Evaluating the traces of $M_{jk}(x)$, $M^2_{jk}(x)$ and $M^3_{jk}(x)$ we find
\bea
&& \Lambda_1(x)+\Lambda_2(x)+\Lambda_3(x)=0, \nonumber \\
&& \Lambda^2_1(x)+\Lambda^2_2(x)+\Lambda^2_3(x)=\frac{E^a(x)E^a(x)}{2}, \nonumber \\
&& \Lambda^3_1(x)+ \Lambda^3_2(x) +\Lambda^3_3(x) =\frac{1}{4} [d_{abc}E^a(x)E^b(x)E^c(x)]
\label{tr1}
\eea
the solution of which is given by eq. (\ref{lambc}).

Using the eigenvalues of $M_{jk}(x)$
from eq. (\ref{lambc}) we perform the color trace in eq. (\ref{i2ss4}) and find
\bea
&& F_l=\frac{1}{(2\pi)^2} \sum_{j=1}^3 [\int dx  \int dp_x
\int dt' \int dt'' \int dp'_0  e^{-it' p'_0} \nonumber \\
&& <t'|e^{is[ -g^2\Lambda^2_j(x)t^2+\hat{p}_0^2+ig\lambda_l \Lambda_j(x)]}|t''>e^{it'' p'_0}].
\label{i2ss45}
\eea

Since the eigenvalues $\Lambda_j$'s are c-numbers, the above
equation involves one harmonic oscillator
\bea
[\frac{1}{2} \omega^2(x)t^2+\frac{1}{2}{\hat p}_0^2]|n>= (n+\frac{1}{2})
\omega(x)|n>
\eea
with space-dependent frequency $\omega(x)$. The harmonic
oscillator wave function
\bea
<t|n>=\psi_n(t)=(\frac{\omega(x)}{\pi})^{1/4}\frac{1}{(2^nn!)^{1/2}}
H_n(t \sqrt{\omega(x)})e^{-\frac{\omega(x)}{2}t^2}
\label{hmx}
\eea
with space dependent frequency $\omega(x)$ is normalized,
\bea
\int dt |<t|n>|^2=1,
\label{norm}
\eea
where $H_n$ is the Hermite polynomial.
Inserting complete set of harmonic oscillator states
(by using $\sum_n |n><n|=1$) in eq. (\ref{i2ss45}) we find
 \bea
&& F_l
=\frac{1}{(2\pi)^2} \sum_{j=1}^3 \sum_n \int dx  \int dp_x \int dt' \int dt'' \int dp'_0 e^{-it' p'_0}
<t'|n>e^{-s[g\Lambda_j(x)(2n+1)+g\lambda_l \Lambda_j(x)]} \nonumber \\
&& <n|t''>e^{it'' p'_0}
=\frac{1}{2\pi} \sum_{j=1}^3 \int dx  \int dp_x \int dt~|<t|n>|^2~e^{-sg\lambda_l \Lambda_j(x)} \frac{1}{2~ {\rm sinh} (sg\Lambda_j(x))} \nonumber \\
&& = \frac{1}{4\pi} {\rm tr}_{\rm color} [ \int dx  \int dp_x \frac{e^{-sg\lambda_l M(x)}}{{\rm sinh}(sgM(x))}]_{jk}.
\label{fn2}
\eea
where we have used eq. (\ref{norm}).
The Lorentz force equation: $\delta_{jk} dp^\mu~=~gT^a_{jk}F^{a \mu \nu}(x) dx_\nu$, of the quark in color space
becomes (when $E^a(x)$ is along the $x$-axis, eq. (\ref{7c}))
\bea
\delta_{jk}dp_x=gT^a_{jk}E^a(x) dt = gM_{jk}(x) dt.
\label{lore}
\eea
Implementing this in eq. (\ref{fn2}) we find
\bea
F_l
=\frac{1}{4\pi} \sum_{j=1}^3 \int dx  \int dt ~\frac{g\Lambda_j(x) e^{-sg\lambda_l \Lambda_j(x)}}{{\rm sinh}(sg\Lambda_j(x))}.
\label{fnn}
\eea

\subsection{ The Generating Functional }

Using the above expression of $F_l$ in eq. (\ref{15cs}) we obtain
\bea
 W= \frac{i}{32 \pi^3} \sum_{l=1}^4 \sum_{j=1}^3
\int_0^\infty \frac{ds}{s} \int dt \int d^3x \int d^2p_T
e^{-is(p_T^2+m^2-i\epsilon)}
[ \frac{g\Lambda_j(x) e^{-sg\lambda_l \Lambda_j(x)}}{{\rm sinh}(sg\Lambda_j(x))}
-\frac{1}{s} ]
\label{15csf}
\eea
where $p_T^2=p_y^2+p_z^2$. Summing over $l$, by using the
eigenvalues ($\lambda_l$) of the Dirac matrix $\gamma^0 \gamma^1$
from eq. (\ref{eigend}) and integrating over $p_T$ we find
\bea
 W= \frac{1}{8\pi^2}~\sum_{j=1}^3
\int dt \int d^3x \int_0^\infty \frac{ds}{s^2}  e^{-is(m^2-i\epsilon)}
~[ g\Lambda_j(x) ~{\rm coth} (gs \Lambda_j(x))
-\frac{1}{s}
].
\label{15csfa}
\eea
This equation is divergent as $s \rightarrow $ 0.
This ultraviolet divergence can be removed by charge renormalization
\cite{schw} by subtracting also the term linear in $s$ in the expansion of
$\frac{{\rm cosh} (s g\Lambda_j(x))}{{\rm sinh }(sg\Lambda_j(x))}$. Hence
after renormalization we obtain
\bea
 W= \frac{1}{8\pi^2}~\sum_{j=1}^3
\int dt \int d^3x \int_0^\infty \frac{ds}{s^2}  e^{-is(m^2-i\epsilon)}
~[ g\Lambda_j(x)~ {\rm coth} (gs \Lambda_j(x)) -\frac{1}{3}sg^2\Lambda^2_j(x)
-\frac{1}{s} ].
\label{15csfb}
\eea

Using eq. (\ref{15csfb}) in (\ref{gen})
we find the following expression for the generating
functional of quark (antiquark) in the presence of
arbitrary space-dependent static color potential $A^a_0(x)=-\int dx E^a(x)$ with arbitrary
color index $a$=1,2,...8 in SU(3),
\bea
&& \frac{Z[A]}{Z[0]}=\frac{\int [d\bar{\psi }][d\psi ] e^{i\int d^4x \bar{\psi}^j(x)(\delta_{jk}{{\hat p}\!\!\!\slash} ~-gT^a_{jk}{A\!\!\!\slash}^a -\delta_{jk}m) \psi^k(x)}}{\int [d\bar{\psi }][d\psi ] e^{i\int d^4x \bar{\psi}^j(x)({{\hat p}\!\!\!\slash}  ~-m) \psi^k(x)}} =  \nonumber \\
&&\exp \big \{\frac{i }{8\pi^2}~\sum_{j=1}^3 \int dt
\int d^3x \int_0^\infty \frac{ds}{s^2}e^{-is(m^2-i\epsilon)}
~[ g\Lambda_j(x) ~{\rm coth} (gs \Lambda_j(x))
-\frac{1}{3}sg^2\Lambda^2_j(x)  -\frac{1}{s}
]\big \} \nonumber \\
\label{fin}
\eea
which reproduces eq. (\ref{n}). The gauge invariant expressions for $\Lambda_j(x)$'s are given
in eq. (\ref{lambc}).

\section{Conclusion}

To conclude we have performed the path integral
for a quark (antiquark) in the presence of arbitrary
space-dependent static color potential $A^a_0(x)(=-\int dx E^a(x)$)
with arbitrary color index $a$=1,2,...8. We have shown that such a path integration is possible even if one
can not solve the Dirac equation in the presence of arbitrary space-dependent
potential. In particular we have obtained an
exact expression of the non-perturbative generating functional of
quark (antiquark) in the presence of arbitrary space-dependent color potential
$A^a_0(x)(=-\int dx E^a(x)$) with arbitrary color index $a$=1,2,...8 in SU(3).
This result crucially depends on the validity of the shift conjecture which has
not yet been established.

It may be possible to further explore this path integral technique
to study non-perturbative bound state formation, for example, by studying
various non-perturbative correlation functions.

\acknowledgements
I thank Robert Shrock, Jack Smith and George Sterman for useful
discussions. I also thank Jack Smith and George Sterman
for careful reading of the manuscript.
This work was supported in part by the National
Science Foundation, grants PHY-0354776 and PHY-0345822.


\begin{thebibliography}{99}

\bibitem{wilson} K. G. Wilson, Phys. Rev. D10 (1974) 2445.
\bibitem{latt} H. J. Rothe, "Lattice Gauge Theories",  {\it world scientific lecture
notes in physics}-vol. 74, (2005); and references therein.
\bibitem{latt1} J. D. Stack, Phys. Rev. D27 (1983) 412; Phys. Rev. D29 (1984) 1213;
H.-Q. Ding, Phys. Rev. D42 (1990) 2350;
M. Baker, J. S. Ball and F. Zachariasen, Phys. Rev. D56 (1997) 4400; and
and references therein.
\bibitem{shift} F. Cooper and G. C. Nayak, hep-th/0609192.
\bibitem{nayak1} G. C. Nayak, Phys. Rev. D72 (2005) 125010;
G. C. Nayak and P. van Nieuwenhuizen, Phys. Rev. D 71 (2005) 125001;
F. Cooper and G. C. Nayak, Phys. Rev. D 73 (2006) 065005.
\bibitem{nayak2} F. Cooper and G. C. Nayak, hep-th/0611125; hep-th/0612292.
\bibitem{schw} J. Schwinger, Phys. Rev. 82 (1951) 664.

\end{thebibliography}
\end{document}